\documentstyle[12pt]{article}
\newcommand{\be}{\begin{equation}}
\newcommand{\ee}{\end{equation}}
\newcommand{\bea}{\begin{eqnarray}}
\newcommand{\eea}{\end{eqnarray}}
\newcommand{\beas}{\begin{eqnarray*}}
\newcommand{\eeas}{\end{eqnarray*}}
\newcommand{\bi}{\begin{itemize}}
\newcommand{\ei}{\end{itemize}}
\newcommand{\bc}{\begin{center}}
\newcommand{\ec}{\end{center}}
\newcommand{\bfl}{\begin{flushleft}}
\newcommand{\efl}{\end{flushleft}}
\newcommand{\bfr}{\begin{flushright}}
\newcommand{\efr}{\end{flushright}}
\newcommand{\f}{\frac}

\newcommand{\au}{\u{a}}

\newcommand{\s}{\c{s}}

\newcommand {\equ}[1] {(\ref{#1})}

\def\6{\partial} \def\a{\alpha} \def\b{\beta}
 \def\d{\delta} \def\ve{\varepsilon}
\def\e{\epsilon}
  
 \def\k{\kappa} 
\def\m{\mu} \def\n{\nu} \def\x{\xi} 
 \def\s{\sigma} 
\def\Ph{\phi} \def\ph{\varphi} 
\def\o{\omega}  \def\D{\Delta}
 \def\L{\Lambda}

\def\non{\nonumber\\}
\def\={\!\!\!&=&\!\!\!}
\def\+{\!\!\!&&\!\!\!+~}
\def\-{\!\!\!&&\!\!\!-~}

\def\hp{h_{++}}
\def\hm{h_{--}}
\def\hpm{h_{\pm \pm}}
\def\cp{c^{+}}
\def\cm{c^{-}}
\def\cpm{c^{\pm}}
\def\czp{c^{0}_{+}}
\def\czm{c^{0}_{-}}
\def\cup{c^{1}_{+}}
\def\cum{c^{1}_{-}}
\def\xzu{X^{0,1}}
\def\xuz{X^{1,0}}
\def\xuu{X^{1,1}}
\def\Dp{\Delta_{+}}
\def\Dm{\Delta_{-}}
\def\Dpm{\Delta_{\pm}}

\def\hm{h_{--}}
\def\hp{h_{++}}
\def\hpm{h_{\pm\pm}}
\def\w{dx^{+}\wedge dx^{-}}

\def\iup{\xi _{+}^{1}}
\def\ium{\xi _{-}^{1}}
\def\izp{\xi _{+}^{0}}
\def\izm{\xi _{-}^{0}}

\renewcommand{\AA}{{\cal A}}

\newcommand{\CC}{{\cal C}}

\newcommand{\MM}{{\cal M}}

\begin{document}
\title{BRST COHOMOLOGY FOR 2D GRAVITY}
\author{{\em Paul A. Blaga,} \\
Department of Geometry \\
{\em Liviu T\u{a}taru and Ion V. Vancea,}  \\
Department of Theoretical
Physics,\\  Babe\c{s} - Bolyai University of Cluj,\\ Romania}
\maketitle
\begin{abstract}
The BRST cohomology group in the space of local functionals of the
fields for the two-dimensional conformally invariant gravity is
calculated. All classical local actions (ghost number equal to
zero) and all candidate anomalies are given and discussed for our
model.
\end{abstract}
\setcounter{page}{0}
\thispagestyle{empty}

\newpage
\section{Introduction}

Gauge fields play a very important role in all theories which
describe  the fundamental interactions \cite{YM}. The most
efficient way to study the quantization and the renormalization of
all gauge (local) theories is given by the $BRST$ transformations
and by the so-called $BRST$ cohomology \cite{HT,BV,BBH}. Despite
the fact that gravity could be introduced as a gauge  theory
associated with local Lorentz invariance \cite{Uti}, its action has
a  different structure and it is difficult to connect it to a
special  form of the Yang-Mills theory, known as the topological
quantum field theory \cite{Wi}. However, in the $BRST$ quantization
framework the structure of the  invariant action, the anomalies and
the Schwinger terms can be obtained in a  purely algebraic way by
solving the $BRST$ consistency conditions in the space  of the
integrated local field polynomials \cite{St,D}. This fact has been
known since the work of Wess and Zumino \cite{WZ} and for a general
gauge theory the general form of these can be elegantly formulated
in the BV formalism  \cite{BV}. In this framework, the Wess-Zumino
consistency condition \cite{St,D} can be written as:
\be
sA = (S,A) = 0    \label{ceq}
\ee
where $S$ is the proper solution of the master equation, $A$ are
integrated  local functionals and $s$ is the {BRST} nilpotent
differential \cite{St,D,BRST}. The   solutions of ~(\ref{ceq})
modulo the exact forms $A = sB$ can be organized in an abelian
group $H(s)$, the BRST cohomology group. We can introduce a
graduation in  $H(s)$ by the ghost number ($gh = g$) and we can
decompose it in a direct sum  of subspaces with a definite ghost
number:
\be
H(s) = \oplus H^{g}(s)
\ee
Anomalies are represented by cohomology classes of $H^{1}(s)$, but
it is often  useful to compute $H^{g}(s)$ for other values of g as
well since $H^{0}$  contains the
$BRST$ invariant action and $H^{2}(s)$the Schwinger terms. Besides,
the whole $H(s)$ could play an important role in solving and
understanding the descent equations \cite{St,D} and in the study of
structure of the field configurations.\\ In this paper we are going
to investigate the structure of $H(s)$ for a class  of two
dimensional models which are {\bf conformally invariant} at the
classical  level. We will not characterize these models by specific
conformally invariant  classical actions but we rather specify the
field content and the gauge  invariances of the theory. We will try
to obtain $H(s)$ for a general  framework independent on the local
classical action $S_{0}$. Thus we are not going to introduce the
antifields \cite{BV,HT} which have the BRST transformations
dependent on $S_{0}$.\\
The paper is organized as follows: in Sect.2 we recall the field
content and the  gauge symmetries of our model. In Sect.3 we give
the equations which have to be  solved in order to find out
invariant Lagrangians, anomalies and  Schwinger terms. Sect.4 deals
with the analysis of the algebra of all fields and their
derivatives. We shall show that $A$ can be splitted in its
contractive part  $C$ and its minimal subalgebra $M$. In the
minimal algebra we introduce,   following \cite{BTP} a very
convenient basis where the solution of ~(\ref{ceq}) are very
simple. At the end we give the structure of {\bf all} nontrivial
sectors of $H(s)$ and in Sect.5 we give some comments of our
results.

\section{Two-dimensional conformal gravity}

Our main aim is to compute the $BRST$ cohomology group for a class
of  two-dimensional models which are conformally invariant at the
classical level. We are not going to characterize these models by
the local classical action  $S_{0}$, but we will provide only the
field contents and the gauge invariance of the classical theory.
Then we can define the $BRST$ differential $s$ and we can compute
the $BRST$ cohomology group $H^*(s)$ in the space of local
functionals of the fields. Thus, for ghost number zero, this group
provides  the most general local classical action $S_0$. For ghost
number one, $H^1(s)$ gives us the most general local anomalies. In
this paper we are not going to take into account the antifields and
we do not need the concrete form of the classical action $S_0$
which enters the $BRST$  transformations.

The fields of our theory are the components of the $2D$ metric
$g_{\m \n} = g_{\n \m}$ and the set of bosonic scalar matter fields
($X=X^A, A=1,\ldots,D$). It is convenient to replace the metric
$g_{\m\n}$ by the zweibein fields $e^{a}_{\m}$ such that:
\be
g_{\m\n}=e^{a}_{\m}e^{b}_{\n} \eta_{ab}  \label{met}
\ee
where $\m,\n = 0,1$, $a,b = 0,1$ and $\eta_{ab}=(+,-)$ or by the
moving frame
\be
e^a = e^{a}_{\m}dx^{\m}   \label{mov}
\ee
The conformal properties of the two-dimensional spacetime are most
conveniently described in terms of light-cone coordinates:
$$x^{\pm}=\f{1}{\sqrt{2}}(x^0 \pm x^1)$$ and the differential
operators
\be
\f{\6}{\6 x^{\pm}}= \f{1}{\sqrt{2}}\left( \f{\6}{\6 x^0} \pm
\f{\6}{\6 x^1}\right) \label{der}
\ee
For the moving frame one defines the one-forms:
\be
e^{\pm} = \f{1}{\sqrt{2}}(e^0 \pm e^1) \label{mov1}
\ee
with the coefficients
\be
e^+ = (dx^{+} + \hm dx^{-})e^{+}_{+}   \label{coef+}
\ee
\be
e^{-} = (dx^{-} + \hp dx^{+})e^{-}_{-}   \label{coef-}
\ee
Here
\be
\hp = \f{e^{+}_{-}}{e^{+}_{+}} \hspace{1cm} \hm=\f{e^{-}_{+}}{e^{-
}_{-}} \label{gauge}
\ee
are the gauge fields which occur for the nonchiral Virasoro
algebra.  They can be expressed in terms of the components of the
metric $g_{\m \n}$ by using the definitions ~(\ref{met}) and
{}~(\ref{mov}). The metric becomes: \be
ds^2 = g_{\a \b}dx^{\a}dx^{\b}  \label{dist1}
\ee
where $\a,\b = +,-$. The components of the metric in light-cone
frame read out  as:
\begin{eqnarray}
g_{\pm \pm} &=& \f{1}{2}\left(g_{00} \pm 2g_{01} +g_{11}\right)
\nonumber \\ g_{+-} &=&  \f{1}{2}\left(g_{00}-g_{11} \right)  \\
g_{-+} &=& g_{+-} \nonumber \label{met1}
\end{eqnarray}
 and employing the zweibein
\be
ds^2 = \eta_{ab}e^a e^b   \label{dist2}
\ee
Thus, we can write the gauge fields as follows:

\be
\hpm = \f{g_{\pm \pm}}{g_{+-} + \sqrt{g}}   \label{gauge1}
\ee
with $g =|det(g_{\m \n})|$.

The fields $\hpm$ are inert under local Lorentz (structure) group
transformations and change under diffeomorphisms (i.e., from one
chart to another) in the following way:
\be
\hp = \f{(\6x'^{+}/\6x^-) + (\6x'^{-}/\6x^-)\hp'(x')}
{(\6x'^{+}/\6x^+) + (\6x'^{-}/\6x^+)\hp'(x')}  \label{gauge2} \ee
\be
\hm = \f{(\6x'^{-}/\6x^+) + (\6x'^{+}/\6x^+)\hm'(x')}
{(\6x'^{-}/\6x^-) + (\6x'^{+}/\6x^-)\hm'(x')}  \label{gauge3} \ee

Besides, the fields $\hpm$ remain invariant under the Weyl
transformation $g_{\a\b} \rightarrow e^{\s}g_{\a\b}$. The conformal
factors $e^{+}_{+}$ and $e^{-}_{-}$ carry entirely the local
Lorentz group transformations and the Weyl transformation. Thus, if
we want to define a theory that is invariant under the local Weyl
transformations of the metric $g_{\a\b}$ we have to work only with
the fields $\hpm$ instead of the whole metric $g_{\a\b}$

In the parametrization ~(\ref{coef+}) and ~(\ref{coef-}) of the
moving frame, the structure group and Weyl transformations are
{\em carried entirely} by the conformal factors $\e^{\pm}_{\pm}$.
Besides, upon general coordinate transformations they change as
\be
e_+^+(x)=\left[\f{\6  x'^+}{\6 x^+}+\f{\6 x'^-}{ \6
x^+}\hp'(x)\right]e'^+_+(x') \label{conf+}
\ee
and
\be
e_-^-(x)=\left[\f{\6 x'^-}{\6 x^-}+\f{\6 x'^+}{ \6 x^-
}\hm'(x)\right]e'^-_-(x') \label{conf-}
\ee
 Under a Weyl transformation, the fields $\hpm$ and the matter
fields $X$ remain invariant and the conformal factors $e^{+}_{+}$
and $e^{-}_{-}$ change as:
\begin{eqnarray}
e^{+}_{+}(x) &=& e^{-\f{\s}{2}}e'^{+}_{+}(x)\\
e^{-}_{-}(x) &=& e^{-\f{\s}{2}}e'^{-}_{-}(x) \nonumber
\label{conf} \end{eqnarray}
Upon a general coordinate transformation they change as:
\be
e^{+}_{+}(x) = [\frac{\6 x^{'+}}{\6 x^{+}} + \frac{\6 x^{'-}}{\6
x^{+}} h^{'}_{++}(x)]e^{'+}_{+}(x)
\ee
\be
e^{-}_{-}(x) = [\frac{\6 x^{'-}}{\6 x^{-}} + \frac{\6 x^{'+}}{\6
x^{-}} h^{,}_{--}(x)]e^{'-}_{-}(x)
\ee
The geometric description, developed so far, provides the
base for the description of the two-dimensional theories which have
local reparametrization ( diffeomorphism) as
well as Weyl and Lorentz invariance. Upon quantization, the whole
theory must undergo the standard BRST treatment ( or quantization)
to control the degeneracies due to the local gauge transformations.
This will be done in the next section.

The matter field $X=\{X^\m\}$ are scalar under the diffeomorphism,
i.e.,  they have the following transformation under a general
coordinate  transformation:
\be
X(x)=X'(x'). \label{ext}
\ee
We will suppose that $X(x)$ have the Weyl weight zero, i.e., they
are invariant  under the Weyl transformation. In fact, in D
dimensions the Weyl weight  for $X(x)$ is
$$
w(X)=-\f{D-2}{2} $$ which is zero for $D=2$.
\section{BRST transformations of the model}
\setcounter{equation}{0}
The $BRST$ transformations and the $BRST$ differential can be
obtained in general by simply replacing the infinitesimal
gauge parameters $\ve^\a$ which occur in the gauge transformation:
$$\d_{\ve}\ph^i = R^{i}_{\a}\ve^{\a}$$
by the ghosts $c^{\a}(x)$:
$$s\ph^i = R^{i}_{\a}c^{\a}$$
The $BRST$ transformations of the ghosts are defined by
demanding that $s$ should be nilpotent.

For an infinitesimal diffeomorphism we have
\be
x'^{\m} = x^{\m} + \ve^{\m}(x) \label{dif}
\ee
with $\ve^0$ and $\ve^1$ infinitesimal arbitrary functions of $x$.
Under the transformation ~(\ref{dif}) $\hpm$ change as:
\begin{eqnarray}
\d\hp &=& \hp(\6_{-}\ve{-} +\hp\6_{-}\ve^{+}) -(\6_{+}\ve^{-} +
\hp\6_{+}\ve^{+}) - (\ve^{+}\6_{+} + \ve^{-}\6_{-})\hp  \\
\d\hm &=& \hm(\6_{+}\ve{+} +\hm\6_{+}\ve^{-}) -(\6_{-}\ve^{+} +
\hm\6_{-}\ve^{-}) - (\ve^{-}\6_{-} + \ve^{+}\6_{+})\hm \nonumber
\label{vgauge}
\end{eqnarray}
where $\6_{+} = \6/\6x^+$ and $\6_{-} = \6/\6x^-$.

The matter field $X$ has the following transformation under the
diffeomorphism ~(\ref{dif}):
\be
\d X = (\ve^{+}\6_{+} + \ve^{-}\6_{-})X    \label{vmat}
\ee

{}From~(\ref{dif}),~(\ref{vgauge}),~(\ref{vmat}) we get the $BRST$
transformations by
simply replacing the infinitesimal gauge parameters $\ve^{\pm}$ by
{\em the  diffeomorphism ghost $\x^{\pm}$.} These transformations
can be simplified if  new ghost
fields $c^{\pm}$ are introduced following the line of
\cite{BTP,B,BB,G}: \be
c^{\pm} = \x^{\pm} + h_{\mp \mp}\x^{\mp}        \label{nghost} \ee

Thus, the $BRST$ transformations of $\hp,\, \hm,\, X$ are:
\begin{eqnarray}
s\hp &=& \6_{+}c^{-} - \hp\6_{-}c^{-} + c^{-}\6_{-}\hp \nonumber \\
s\hm &=& \6_{-}c^{+} - \hm\6_{+}c^{+} + c^{+}\6_{+}\hm \\
sX &=& c^{+}D_{+}X +c^{-}D_{-}X \nonumber
\label{brst}
\end{eqnarray}
where:
\bea
D_{\pm}X &=& \f{1}{1-y}(\6_{\pm}-\hpm\6_{\mp})X = \f{1}{1-y}\nabla _{\pm}X \\
y &=& \hp\hm \nonumber
\label{dif1}
\eea
The $BRST$ transformations of the ghosts $c^{\pm}$ can be obtained
from the  nilpotency of $s$ and the fact that it commutes with
$\6_{\pm}$. In this way  we  conclude that
\be
sc^{\pm} = c^{\pm}\6_{\pm}c^{\pm}       \label{tghost}
\ee

Under the Weyl transformation $\hp$ and $\hm$ remain invariant and
only the conformal factors $e^{+}_{+}$ and $e^{-}_{-}$ change.

In the framework of the $BRST$ transformations, the search for
invariant Lagrangian anomalies and Schwinger terms can be done in
a purely algebraic way, by solving the $BRST$ consistency condition
in the space of integrated local polynomials \cite{WZ,WSS,BDK}.
This amounts to study the nontrivial solutions of the equation
\be
s\AA = 0 \label{cons}
\ee
where $\AA$ is an integrated local functional $\AA = \int{d^2f}$.
The condition ~(\ref{cons}) translates into the local descent
equations \cite{S}:
\bea
s\o_2 + d\o_1 &=& 0     \nonumber \\
s\o_1 + d\o_0 &=& 0     \\
s\o_0 &=& 0     \nonumber \label{des}
\eea
where $\o_2$ is a $2$-form with $\AA = \int{\o_2}$ and
$\o_1,\,\o_0$ are {\sl local} $1$-forms, respectively $0$-forms. It
is well known  \cite{WSS,BDK,BSS} that the descent equations
terminate in the bosonic  string or the superstring in Beltrami or
super-Beltrami parametrization always with a nontrivial $0$-form
$\o_0$ and that their integration is trivial:
\be
\o_1 = \d\o_0; \hspace{1cm} \o_2 = \f{1}{2}\d^2\o_0
\label{des1} \ee
where $\d$ is a linear operator, which allows to express the
exterior  derivative $d$ as a BRST commutator
\be
d= -[s,\d]      \label{des2}
\ee
This operator was introduced by Sorella for the Yang-Mills theory
\cite{S} and it was used for solving the descent equations
\equ{des1} for the bosonic string \cite{WSS} and superstring
\cite{BSS} in the  Beltrami and super-Beltrami parametrization. It
is easy to see that, once  the last equation in the tower
\equ{des}, i.e.,
\be
s\o_0=0 \label{des3}
\ee
is solved, the rest of the equations from \equ{des} can be solved
with  the help of the operator $\d$ with the solutions \equ{des1}.
Thus, due to  the operator $\d$, the study of the cohomology
of$s\bmod d$ is  essentially reduced to the study of the local
cohomology of $s$ which, in  turn, can be systematically analyzed
by using different powerful  techniques from the algebraic topology
as the Sullivan and  K\"{u}nneth theorems, the spectral sequences
\cite{D}, etc. Actually, as proven in  \cite{ST} for the Yang-Mills
theory, the solutions obtained by utilizing  the decomposition
\equ{des1} turn out to be completely equivalent to that  based on
{\em the Russian formula}\cite{St}.

The main purpose of our paper is to solve the descent equations
{}~(\ref{des}) in the algebra of the local polynomials of
all fields and their derivatives $\AA$. A basis of this algebra can
be chosen to consist of:
\be
\{\6^{p}_{+}\6^{q}_{-}\psi,\6^{p}_{+}\6^{q}_{-}c^{\pm}\}
\label{basis} \ee
where $\psi=(X,\hp,\hm)$ and $p,q=0,1,2,\ldots$. However, the
$BRST$  transformations of this basis are quite complicated and
contain many terms which can be eliminated in $H(s)$. In the next
section we shall eliminate a part of this basis and introduce a new
basis in which the action of $s$ is quite simple.
***** 
\section{The structure of the fields algebra}
\setcounter{equation}{0}

The calculation of $H(s)$ can be considerable simplified if we take
into account that
$\AA$ as a {\sl free} differential algebra and make use of a very
strong theorem due to Sullivan. A free differential algebra is an
algebra generated by a basis, endowed with a differential. The
Sullivan's theorem tells us the following:\\
\indent {\sl The most general free differential algebra $\AA$ is a
tensor product of a contractible algebra and a minimal one}

A minimal differential algebra $\MM$ with the differential $s$ is
one for which $s\MM \subset \MM^{+}. \MM^{+}$ being the part  of
$\MM$ in positive degree, i.e. $\MM = \CC\oplus\MM^{+}$ and a
contractible differential algebra $\CC$ is one isomorphic to a
tensor product of algebras of the form $\L(x,sx)$.

On the other hand, due to K\"{u}nneth's theorem the cohomology of
$\AA$  is given by the cohomology of its minimal part and we can
say that the contractible part $\CC$ can be neglected in the
calculation of $H(s)$

For our algebra, the construction of $\CC$ and $\MM$ is
straightforward  and we do not need any general method to
accomplish that. In fact,  it is easy to see from the $BRST$
transformations of $\hpm, X, c^{\pm}$  that the generators
\be
\6^{p}_{+}\6^{n}_{-}c^{+}, \6^{p}_{+}\6^{n}_{-}c^{-}    \label{gen}
\ee
with $p=0,1,2,\ldots$ and $n=1,2,\ldots$ can be replaced by: \be
\{\6^{p}_{+}\6^{p}_{-}X, \6^{p}_{+}\6^{p}_{-}\phi,
s\6^{p}_{+}\6^{p}_{-}\phi,\6^{p}_{\pm}c^{\pm}\}
\label{bas1} \ee

The Sullivan decomposition can be easily obtained
from~(\ref{bas1}).  Indeed, the contractible subalgebra is
generated by:
\be
\{\6^{p}_{+}\6^{p}_{-}\hpm,s(\6^{p}_{+}\6^{p}_{-}\hpm)\}
\label{cgen} \ee
and the minimal subalgebra $\MM$ by:
\be
\{\6^{p}_{+}\6^{p}_{-}X, \6^{p}_{\pm}c^{\pm}\}
\label{mgen} \ee

Now, to calculate the cohomology $H(s)$ we take into account  only
the basis~(\ref{mgen}). Nevertheless, this basis is not convenient
since the differential $s$ has a complicated action on it. The
investigation of the $BRST$ cohomology, i.e., the solution of
{}~(\ref{des3}) is considerable  simplified by using an appropriate
new basis substituting the fields$(X, c^{\pm})$ and their
derivatives. The construction of this new basis is a crucial step
towards the calculation of $H(s)$. This new basis has been proposed
by  Brandt, Troost and Van Proeyen \cite{BTP} and it is basically
intended to substitute one by one the elements of the basis
{}~(\ref{mgen}) by: \be
\D^{p}_{+}\D^{q}_{-}X =X^{p,q}  \label{nbas1}
\ee
\be
\f{1}{(p+1)!}\D^{p+1}_{\pm}\cpm = c_{\pm}^{p}
\label{nbas2} \ee
where
\be
\D_{\pm} = \left\{ s,\f{\6}{\6 c^{\pm}} \right\} =
s\f{\6}{\6 c^{\pm}} + \f{\6}{\6 c^{\pm}}s .      \label{nbas3} \ee

It is crucial to remark that the linear operators $\Dpm$ act on
the algebra $\AA$ as derivatives, i.e., they obey the Leibnitz
rule:
\be
\Dpm(ab)= \Dpm a\cdot b + a\Dpm b       \label{leib}
\ee
The action of $s$ on the elements ~(\ref{nbas1}) which form a new
basis, can be calculated directly
\be
sX^{p,q} =\Dp^p \Dm^q (sX) =
\sum_{k=0}^{\infty}\left[ \left(\begin{array}{c}
p\\k\end{array} \right)
(\Dp^k \cp)X^{k+1,q} + \left(\begin{array}{c}
q\\k\end{array} \right) (\Dm^k \cm)X^{p,k+1} \right]
\label{brf} \ee
\be
sc^{n}_{\pm} = \f{1}{(n+1)!}\D^{n+1}_{\pm}(c^{\pm}\D_{\pm}c^{\pm})
\label{sc} \ee
since $s$ commutes with $\D_{\pm}$ and
$$ sX=\cp\xuz +\cm\xzu $$ $$ \D_{\pm}c^{\mp}=0. $$
The remarkable property of this new basis is the fact that its
$BRST$  transformation is given by the Virasoro algebra with
associated ghosts just $c^{\pm}$. Indeed, the last two equations
{}~(\ref{brf}),~(\ref{sc}) can be rewritten as:
\be
sX^{p,q} = \sum _{k\geq -1}(c^{k}_{+}L^{+}_{k} + c^{k}_{-}L^{-
}_{k})X^{p,q}  \label{vir} \ee
\be
sc^{k}_{\pm} = {\f{1}{2}}f_{mn}^{k}c^{m}_{\pm}c^{n}_{\pm}
\label{vg}  \ee
where $L^{+}_{k}$ and $L^{-}_{k}$ are given by the following
equations: \be
L^{+}_{k}X^{p,q} = A^{p}_{k}X^{p-k,q}       \label{vg1}
\ee
\be
L^{-}_{k}X^{p,q} = A^{p}_{k}X^{p,q-k}       \label{vg2}
\ee
and
\be
A^{p}_{k} = \f{p!}{(p-k-1)!}.
\ee
The $ BRST $
transformations  of the ghosts $c^{p}_{\pm}$ can easily be written
as:
\be
sc^{p}_{\pm}=\f{1}{(n+1)!}\D_{\pm}^{n+1}(c^{\pm}\6_{\pm}c^{\pm}) =
\sum_{k=-1}^{p}(p-k)c_{\pm}^{k}c_{\pm}^{p-k} = \f{1}{2}
\sum_{m,n\geq0}f^{p}_{m,n}c_{\pm}^m c_{\pm}^n   \label{brs} \ee
where
\be
f^{p}_{m,n} = (m-n)\delta^{p}_{m+n}   \label{fmn}
\ee
are the structure constants of the Virasoro algebra.

Now, it is easy to see that $L^{\pm}_n$ represent, on the basis
$X^{p,q}$, the  Virasoro algebra according to:
\be
[L^{\pm}_m,L^{\pm}_n] = f_{m,n}^k L^{\pm}_k \hspace{1cm}
[L^{+}_m, L^{-}_n] = 0 \label{vir1}
\ee

{}From~(\ref{vir}) we can give another representation for the
generators$L^{\pm}_n$ on the algebra spanned by $X^{p,q}$. They
have the form:
\be
L^{\pm}_n =  \left\{s,\f{\6}{\6 c^{n}_{\pm}} \right\} \label{vir2}
\ee
and they can be extended to $c^{n}_{\pm}$.\\
It is worthwhile remarking that on the basis
${c^{p}_{\pm},X^{p,q}}$ the  BRST differential $s$ can be written
in the form:
\be
s = \sum_{k=1}(c^{k}_{+}L^{+}_{k} + c^{k}_{-}L^{-}_{k}) +
\frac{1}{2}\sum_{m,n,k}f^{k}_{m,n}(c_{+}^{m}c_{+}^{n}\frac{\partial}
{\partial c^{k}_{+}}
+ c^{m}_{-}c^{n}_{-}\frac{\partial }{\partial c^{k}_{-}})
\label{es} \ee
The generators $L^{\pm }_{0}$ are {\bf diagonal} on all elements of
the new basis. Indeed, one has:
\be
L^{+}_{0}X^{p,q} = pX^{p,q} \hspace{0.5cm} L^{-}_{0}X^{p,q} =
qX^{p,q} \label{pve} \ee
\be
L^{\pm}_{0}c_{\pm}^{p} = pc_{\pm}^{p} \label{pvc}
\ee
\be
L^{\pm}_{0}c_{\mp}^{p} = 0  \label{pvz}
\ee
Thus, any product of elements of this basis is an eigenvector of
the $L^{\pm}_{0}$.\\ Due to the fact that $L^{\pm}_{0}$ have the
form:
\be
L^{\pm}_{0} = s\frac{\partial}{\partial c^{\pm}_{0} } +
\frac{\partial}{\partial c^{\pm}_{0}} \label{ls} \ee
we can conclude that the solutions of $s\omega _{0} = 0$ must have
the total weight $(0,0)$, all other contributions to $\omega _{0}$
being trivial.\\ All monomials with weight $(0,0)$ are tabled
below:
\bc
\begin{tabular}{||c|c|c||} \hline\hline
$Ghost$   &  $Monomial$    & $s(Monomial)$              \\\hline\hline
$0$     &  $F$             & $(c^{+}X^{1,0}+ c^{-}X^{0,1})\6 F$ \\\hline
$1$     &  $c^{0}_{+}F$    & $2\cp\cup F
                             + c^{0}_{+}\6F(c^{+}X^{1,0}+c^{-}X^{0,1})$ \\
        &  $c^{0}_{-}F$    & $2\cm\cum F + c^{0}_{-}\6F(c^{+}X^{1,0}+c^{-}
                             X^{0,1})$ \\
        &  $c^{+}X^{1,0}F$ & $-\cp\cm\xuz\xzu\6 F + \cp\cm\xuu F$ \\
        &  $\cm\xzu F$     & $-\cp\cm\xuz\xzu\6 F - \cp\cm\xuu F$ \\ \hline
$2$     &  $\cp\cup F $    & $-\cp\cm\cup\xzu\6 F$ \\
        &  $\cm\cum F $    & $\cp\cm\cum\xuz\6 F$  \\ \hline
\end{tabular}
\ec
\bc
\begin{tabular}{||c|c|c||} \hline\hline
$Ghost$   &  $Monomial$    & $s(Monomial)$              \\
\hline\hline
$2$     &  $\czp\czm F$    & $\czp\czm(\cp\xuz + \cm\xzu)\6 F +
                             2(-\czp\cm\cum +\cp\cup\czm) F$\\
        &  $\cp\cm\xuu F$  & $0$ \\
        &  $\cp\cm\xuz\xzu F$ & $0$ \\
        &  $\cp\czp\xuz F$ & $-\cp\cm\czp\xuu F + \cp\cm\czp\xuz\xzu\6F$\\
        &  $\cm\czp\xzu F$ & $\cm\czm\czp\xzu F - 2\cm\cp\cup\xzu F + $\\
        &                  & $ \cm\czp\czm\xzu F + \cp\cm\czp\xuz\xzu\6F$\\
        &  $\cm\czm\xzu F$ & $-\cp\cm\czm\xzu\xuz\6 F -\cp\cm\czm\xuu F$\\
        &  $\cp\czm\xuz F$ & $-2\cp\cm\cum\xuz F -
                             \cp\cm\czm\xuz\xzu\6 F + \cp\cm\czm\xuu\ F$ \\
                             \hline
$3$     &  $\cp\cup\czp F$ & $\cp\cup\czp\cm\xzu\6 F$ \\
     &  $\cm\cum\czm F$ & $-\cm\cum\czm\cp\xuz\6 F$ \\
     &  $\cp\cup\czm F$ & $-\cp\cup\czm\cm\xzu\6 F +2\cp\cup\cm\cum
F$ \\     &  $\cm\cum\czp F$ & $-\cm\cum\czp\cp\xuz\6 F
+2\cm\cum\cp\cup F$ \\   &  $\czp\czm\cp\xuz F$ & $-
\czp\cp\cm(\czm\xuz\xzu\6 F + \czm\xuu F +
     2 \cum\xuz F)$ \\
     &  $\czp\czm\cm\xzu F$ & $-\czm\cm\cp(\czp\xzu\xuz\6 F +
\czp\xuu F +                              2 \cup\xzu F)$ \\
     &  $\cp\cm\czp\xuz\xzu F$ & $0$ \\
     &  $\cp\cm\czm\xuz\xzu F$ & $0$ \\
     &  $\cp\cm\czp\xuu F$ & $0$ \\
     &  $\cp\cm\czm\xuu F$ & $0$ \\
     &  $\cp\cup\cm\xzu F$ & $0$ \\
     &  $\cm\cum\cp\xuz F$ & $0$ \\ \hline
$4$     &  $\cp\cup\czp\czm F$ & $\cp\cup\czp\czm\cm\xzu\6 F -
                      2 \cp\cup\czp\cm\cum F$ \\
     &  $\cm\cum\czp\czm F$ & $\cm\cum\czp\czm\cp\xuz\6 F +
                      2 \cm\cum\czm\cp\cup F$ \\
     &  $\cp\cm\czp\czm\xuz\xzu F$ & $0$ \\
     &  $\cp\cm\czp\czm\xuu F$ & $0$ \\
     &  $\cp\cm\czp\czm\xzu F$ & $0$ \\
     &  $\cp\cm\czm\cup\xzu F$ & $0$ \\
     &  $\cp\cm\czp\cup\xzu F$ & $0$ \\
     &  $\cp\cm\czm\cum\xuz F$ & $0$ \\
     &  $\cp\cum\cm\cum F$     & $0$ \\ \hline
\end{tabular}
\ec

\bc
\begin{tabular}{||c|c|c||} \hline\hline
$Ghost$   &  $Monomial$    & $s(Monomial)$              \\
\hline\hline
$5$     &  $\czp\czm\cp\cup\cm\xzu F$ & $0$ \\
     &  $\czp\czm\cm\cum\cp\xuz F$ & $0$ \\
     &  $\czp\cp\cup\cm\cum F$ & $0$ \\
     &  $\czm\cp\cup\cm\cum F$ & $0$ \\ \hline
$6$     &  $\cp\cup\cm\cum\czp\czm$ & $0$ \\ \hline\hline
\end{tabular}
\ec
\bc
TABLE 1
\ec
where $F=F(X)$ is an arbitrary smooth function of the
matter fields $X$ and  $\6 F = \6 F/\6 X$.\\
{}From this table we can calculate the cohomology group $H^{g}(s)$ by
finding out which of the combinations of the above monomials are
non trivial solutions of the equation $s\omega _{0} = 0$.\\ The
independent non trivial solutions are given in Table 2.
\bc
\begin{tabular}{||c|c||} \hline\hline
$Ghost$ &  $Monomial$ \\ \hline\hline
  $0$   &              -           \\   \hline
  $1$   &              -          \\     \hline
  $2$   &  $\cp\cm\xuz\xzu F$           \\ \hline
  $3$   &  $\cp\cm\czp\xuz\xzu F$  \\
        &  $\cp\cm\czm\xuz\xzu F$  \\
        &  $\cp\cup\czp$  \\
        &  $\cp\cum\czm$ \\ \hline
  $4$   &  $\cp\cm\czp\cum\xuz F$ \\
        &  $\cp\cm\czm\cup\xzu F$  \\
        &  $\cp\cm\czp\czm\xuz\xzu F$ \\ \hline
  $5$   &  $\czp\czm\cp\cup\cm\xzu F$ \\
        &  $\czm\cp\cup\cm\cum\xuz F$ \\ \hline
  $6$   &  $\cp\cup\cm\cum\czp\czm $ \\ \hline\hline
\end{tabular}
\vspace{0.5cm}
\ec
\bc
TABLE 2
\ec
REMARKS.
\begin{enumerate}
\item In the Table 2 we have included only independent solutions of
Eq.~(\ref{des3}) and we have not given the solutions which differ
by those from Table 2 by an s-exact term. For instance
for ghost $g=2$ there  are  two  additional solutions
of Eq.~(\ref{des3}) given by

\bea
\eta_1=\cp\cm\czp\xuu F  \\
\eta_2=\cp\cm\czm\xuu F
\eea
but it is easy to verify that they are $s$-dependent on the ones
given in  Table 2. Indeed a little algebra  shows that
$$
\eta_1=\cp\cm\czp\xuz\xzu \6 F+s \left[ \cp\czp \xuz F \right] $$
and
$$
\eta_2=\cp\cm\czp\xuz\xzu \6 F+s\left[- \cm\czm \xzu F \right]. $$
\item For gh=4 we have only three independent solutions,
the rest of five being or s-exact or a linear combination of
these three solutions and some s-exact terms. For instance one can
write
\beas
\cp\cm\czp\czm X^{1,1} F&=&-\f1{2}s(\czp\czm X^{1,0}F-\cm\czp\czm
X^{0,1}F )-\cp\cm\czp\cup X^{1,0}F -\cp\cm\czm\cum X^{0,1}F \\ &-&
     \cp\cm\czp\czm\xuz\xzu\6 F
\eeas
and
$$
\cp\cm\czp\cup \xzu \6 F =s(\cp\czp\cup F).
$$
\item For gh=5 we have four solutions but only two of them are
independent.  In this case the non-independent solutions are
\beas
\cp\cm\czp\cup\cum F&=&\f1{2}s(\cp\czp\czm\cup F)+
\cp\cm\czp\czm\cup\xzu F \\
\cp\cm\czm\cup\cum F&=&\f1{2}s(\cm\czp\czm\cum F)+
\cp\cm\czp\czm\cum\xuz F
\eeas
\end{enumerate}

We must point out that all elements of $H(s)$ given in Table 2 are
solutions
in the space of local functions, i.e., in the space of
0-forms. If we want
to compute the $BRST$ cohomology in the space of local functionals
$\omega = \int d^{3}x f $ which fulfil the equation $s\omega = 0$
we have to solve the descent  equation. As we have been pointing
out, these equations can be solved by using the operator $\delta $
defined by
\be
\delta = dx^{\alpha}\frac{\partial}{\6 \xi^{\alpha}} \label{del}
\ee
on this way we can write the integrand of $\omega$,$\omega_{2} =
d^{2}xf$ in the following form
\be
\omega_{2} = \frac{1}{2}\delta\delta\omega_{0}
\ee
and we can compute all terms from the BRST
cohomology in the space of local functionals.

The operator $\d$ can be defined directly on the basis used by us
as \bea
\d \cpm =dx^{\pm}+\hpm dx^{\mp} \non
d\Ph=0 \hspace{0.4cm}\mbox{for}\hspace{0.5cm} \Ph=\{\hpm,X\}  \eea
and in addition
$$
[\d,\6_{\pm}]=0.
$$
Now it is easy to see that $\d$ is of degree zero and obeys the
following  relations
\be
d=-[s,\d] \hspace{2cm}
[d,\d]=0. \label{decom}
\ee
In order to solve the tower \equ{des} we shall make use of the
following  identity
\be
\label{identity}
e^{\d} s=(s+d)e^{\d}
\ee which is a direct consequence of \equ{decom} (see  \cite{ST}).
Thanks to this identity we can obtain the higher cocycles $\o_1$
and  $\o_2$ once a non-trivial solution $\o_0$ is known. Indeed, if
one apply the  identity \equ{identity} to $\o_0$ we get
\be
(s+d)[e^{\d}\o_0]=0. \label{semifin}
\ee
On the other hand one can see from eq. \equ{del} that the operator
$\d$ acts as a translation on the ghosts $\xi^{\pm}$ with the
amount  $dx^{\pm}$ and eq. \equ{semifin} can be rewritten as
\be
(s+d)\o_0(\xi^{\pm}+dx^{\pm},X)=0.
\ee
Thus the expansion of the zero form cocycle
$\o_0(\xi^{\pm}+dx^{\pm},X)$  in power of the one-form $dx^{\pm}$
yields all the cocycles $\o_1$  and $\o_2$.

 In this way we can compute all terms from the $BRST$ cohomology in
the space of local functionals.

Our results are given in Table 3.
\bc
\begin{tabular}{||c|c|c||} \hline\hline
$Ghost$ &     $Monomial$          &      $\delta^{2}(Monomial)/\w$ \\
                                         \hline\hline
$0$  &              -          &       -                \\   \hline
$1$  &              -          &       -                \\   \hline
$2$  & $\cp\cm\xuz\xzu F$     & $2(1-y) \xuz\xzu F$ \\   \hline
$3$  &  $\cp\cm\czp\xuz\xzu F$ & $(1-y)\izp\xuz\xzu F$ \\
     &  $\cp\cm\czm\xuz\xzu F$ & $(1-y)\izm\xuz\xzu F$ \\
     &  $\cp\cup\czp$          & $\6_{+}\cp\6^{2}_{+}\hm -
                                 \6^{2}_{+}\cp\6_{+}\hm $ \\
     &  $\cm\cum\czm$          & $\6_{-}c^{-}\6^{2}_{-}\hp -
                                 \6^{2}_{-}\cm\6_{-}\hp $ \\ \hline
$4$  &  $\cp\cm\czp\cum\xuz F$      & $(1-y)\izm\ium\xuz F $ \\
     &  $\cp\cm\czm\cup\xzu F$      & $(1-y)\izp\iup\xzu F $ \\
     &  $\cp\cm\czp\czm\xuz\xzu F$  & $(1-y)\izm\izp\xuz\xzu F $\\   \hline
$5$  &  $\cp\cm\czp\czm\cup\xzu F$ & $(1-y)\izp\izm\iup\xzu F $ \\
     &  $\cp\cm\czp\czm\cum\xuz F$ & $(1-y)\izp\izm\iup\xuz F $ \\ \hline
$6$  &  $\cp\cm\czp\czm\cup\cum F$ & $(1-y)\izp\izm\iup\ium F $\\ \hline
        \hline
\end{tabular}
\ec
\vspace{0.5cm}
\bc
TABLE 3
\ec
In Table 3 we made use of the following notations
\be
\izp = \6_{+}\xi ^{+} + \hm\6_{+}\xi ^{-}
\ee
\be
\izm = \6_{-}\xi ^{-} + \hp\6_{-}\xi ^{+}
\ee
\be
\iup = \6^{2}_{+}\xi ^{+} + 2\6_{+}\hm ^{-}\6_{+}\xi^{-} +
h_{--}\6^{2}_{+}\xi^{-}
\ee
\be
\ium = \6^{2}_{-}\xi ^{-} + 2\6_{-}\hp\6_{-}\xi^{+} + h_{++}\6^{2}_{-}\xi^{+}
\ee

All solutions of the descent equations  \equ{des} are discussed in
the last section where we will connect our results with the
previous ones. \section{Discussions and Conclusions}

We have determined the complete $BRST$ cohomology group in the
space of local fields and local functionals for a 2D gravitational
theory invariant under diffeomorphisms and local Weyl
transformations.\\
The elements of the $BRST$ group in the space of the local
functionals, i.e.,  the term $\omega _{2}$ in the descent equations
are
particularly interesting because they represent classical actions,
anomalies and Schwinger terms. For $ghost = 0$ there is only one
element of $H^{2}(s)$ and it corresponds to the unique classical
action. This action has the form of the  $\sigma$ - model with a
torsion term. For $ghost = 1$ there are four non-trivial elements
of $H^{3}(s)$ which can be grouped in two types. Representatives of
the first type can be chosen to be  independent of the matter
fields. There are two independent terms of this type: \be
\int d^{2}xc^{\pm}\partial ^{3}_{\pm}h_{\mp \mp}
\ee
and they represent the candidates for the anomalies \cite{B}.
Representatives of the second type depends nontrivially on the
matter fields and have the form: \be
\int d^{2}x(1-y)(\partial _{\pm} \xi ^{\pm} + h_{\pm \pm}\partial
_{\pm} \xi ^{\mp})(D_{+}X)(D_{-}X)  \ee
where
\be
\xi ^{\pm} = \frac {1}{1-y}(c^{\pm}-h_{\pm \pm}c^{\mp})
\ee
are the diffeomorphism ghosts. In fact the anomalies of the second
type cannot  occur in the perturbative calculations since the
classical action $S_{0}$   does not contain
a self-interactive term in the matter fields. Thus, the numerical
coefficient of the corresponding Feynman diagrams automatically
vanishes.

All these solutions have been obtained by Werneck de Oliveira,
Schweda and  Sorella \cite{WSS} and by Brandt, Troost and Van
Proeyen \cite{BTP}.  The remarkable point here is the fact that
these are the only possible solutions  with ghost number less than
two and for the ghost number zero we have  a unique solution.

The solutions of the descent equations \equ{des} with ghost number
bigger than one do not have any direct physical significance.
Nevertheless we will give all of them for the sake of completeness
and for the future use.

For gh=2 we have the following solution of the form:
\be
\AA^2=\int d^2\left[ \xi^0_+\xi^1_-\nabla_+X^\m f^1_\m(X)+
 \xi^0_-\xi^1_+\nabla_-X^\m f^2_\m(X) + \f1{1-y}\xi^{0}_{+} \xi^{0}_{-}
\nabla_{+}X^{\m}
\nabla_{-}X^{\n}f_{\m \n}
\right]
\label{sol2}
\ee
where
$f_{\m}^{1,2}(X), ~~f_{\m \n}(X)$ are some arbitrary functions of $X$.
In this
case the  solutions of eqs.\equ{des} depend only on the
diffeomorphism ghosts $\xi^{\pm}$.

For gh=3 we have also two independent solutions of the form  \be
\AA^3=\int d^2x \xi^0_{+}\xi^0_{-}\left[ \xi^1_{-}(\nabla_-
X^{\m}f^1_{\m}(X)+ \xi^1_{-}(\nabla_+ X^{\m})f^2_{\m}(X)  \right].
\label{sol3} \ee
Again in $\AA^3$ occur only the ghosts $\xi^{\pm}$.

In the last possible case gh=4 we have obtained a unique solution
of the form
\be
\AA^4=\int d^2x (1-y)\xi^0_+\xi^0_-\xi^1_+\xi^1_- F(X)
\label{sol4}
\ee
with $F(X)$ an arbitrary scalar function of $X$.

All solutions with ghost number bigger than one are new and  as far
as we are aware of this is the first place where they are done.

Finally we want to mention that a similar calculation with the
antifields included can be done \cite{BTP}, but in this case  the
results strongly depends on the form of the classical action one
starts with.


\newpage
\begin{thebibliography}{99}
\bibitem{YM}C.N. Yang and R.L. Mills, {\em Phys.Rev.} {\bf
96}(1954) 191; S. Glashow and M. Gell-Mann, {\em Ann. Phys. (N.Y.)}
{\bf 15} (1961) 437; \bibitem{HT} M.Henneaux and C. Teitelboim,
{\em Quantization of Gauge Systems}, Princeton University Press,
1992;
\bibitem{BV} I.A. Batalin and G.A. Vilkovisky, {\em Phys. Rev.}
{\bf D28} (1983) 2567;
\bibitem{BBH} G. Barnich, F. Brandt and M. Henneaux,
preprint ULB-TH-94/06, NIKEF-H 94-13, hep/th/9405109
\bibitem{Uti} R. Utiyama {\em Phys. Rev.} {\bf 101} (1956) 1597;
\bibitem{Wi} E. Witten, {\em Comm. Math. Phys.} {\bf 117}(1988)
353; {\em Comm. Math. Phys.} {\bf 118}(1988) 411;
\bibitem{St} R.  Stora, {\em Algebraic structure and topological
origin of anomalies}, Carg\`{e}se '83,
s. G. t'Hooft et.al., Plenum Press, New York, 1987;
\bibitem{D} J. Dixon, {\em Comm. Math. Phys.} {\bf 139}(1991) 495;
\bibitem{WZ} J. Wess and B. Zumino, {\em Phys. Lett.} {\bf
B37}(1971) 95; B. Zumino,
 {\em Chiral anomalies and differential
geometry}, Les Houches '83, eds. B.S. De Witt and R. Stora, North
Holland, Amsterdam, 1987;
\bibitem{BRST} C. Becchi, A. Rouet and R. Stora, {\em Ann.
Phys.(N.Y.)} {\bf 98}(1976) 287; I.W. Tyutin {\em Gauge Invariance
in Field Theory and Statistical Physics}, Lebedev Institute
preprint FIAN no.{\bf 39}(1975);
\bibitem{BTP} F. Brandt, W. Troost
and A. Van Proeyen ,
NIKHEF-H 94-16, KUL-TF-94/17, hep-th/9407061, to appear in
the proceedings of the {\em Geometry of Constrained Dynamical
Systems} workshop, Isaac Newton Institute for Mathematical
xSciences, Cambridge, June 15-18, 1994; W. Troost and A. Van
Proeyen, KUL-TF-94/94, hep-th/9410162;
\bibitem{WSS} M. Werneck de Oliveira, M. Schweda and S.P.
Sorella, {\em Phys. Lett.} {\bf B315}(1993) 93;
\bibitem{BDK} F. Brandt, N. Dragon and M. Kreuzer, {\em Nucl. Phys}
{\bf B340}(1990) 187;
\bibitem{S} S.P. Sorella {Comm. Math. Phys.}{\bf 157}(1993),231;
\bibitem{ST}S.P. Sorella and L. T\au taru, {\em Phys. Lett.}  {\bf
B324}(1994), 351;
\bibitem{BSS} A. Boresch, M. Schweda and S.P. Sorella, {\em Phys.
Lett.} {\bf B328}(1994);
\bibitem{B} C. Becchi, {\em Nucl. Phys.} {\bf B304} 513;
\bibitem{BBS} L. Baulieu, C. Becchi and R. Stora, {\em Phys. Lett.}
{\bf B180}(1986), 55;
KUL-TF-94/94, hep-th/9410162.
\bibitem{BB} L. Baulieu and M. Bellon, {\em Phys. Lett.} {\bf
B196}(1987), 142; \bibitem{G} R. Grimm {\em Ann. Phys.(N.Y.)} {\bf
200}(1990) 49;

\end{thebibliography}
\end{document}